\documentclass[letterpaper]{article}

\usepackage{amsmath,amssymb,amsthm}
\usepackage[letterpaper,margin=1in]{geometry}

\usepackage{xcolor}
\usepackage{caption}
\usepackage{microtype}
\usepackage{lmodern}
\usepackage{enumitem}
\usepackage{mathtools}
\usepackage{float}
\usepackage{flafter}
\floatstyle{ruled}
\newfloat{algorithm}{htbp}{loa}
\floatname{algorithm}{Algorithm}

\newtheorem{theorem}{Theorem}
\newtheorem{lemma}[theorem]{Lemma}

\usepackage[
    colorlinks=true,
    linkcolor=blue!62!black,
    citecolor=green!48!black,
    urlcolor=blue!70!black
]{hyperref}
\usepackage[nameinlink,noabbrev]{cleveref}
\crefname{theorem}{Theorem}{Theorems}
\Crefname{theorem}{Theorem}{Theorems}
\crefname{lemma}{Lemma}{Lemmas}
\Crefname{lemma}{Lemma}{Lemmas}
\crefname{section}{Section}{Sections}
\Crefname{section}{Section}{Sections}
\crefname{algorithm}{Algorithm}{Algorithms}
\Crefname{algorithm}{Algorithm}{Algorithms}
\newcommand{\proc}[1]{\mbox{\textnormal{\scshape #1}}}

\allowdisplaybreaks

\DeclareMathOperator{\Tr}{Tr}
\DeclareMathOperator{\rank}{rank}
\DeclareMathOperator{\diag}{diag}
\DeclareMathOperator{\im}{im}
\DeclareMathOperator{\sgn}{sgn}
\newcommand{\E}{\mathbb E}
\newcommand{\Prob}{\mathbb P}
\newcommand{\R}{\mathbb R}
\newcommand{\eps}{\varepsilon}
\newcommand{\calA}{\mathcal A}
\newcommand{\Map}{R}
\hypersetup{pdftitle={Linear-Size Spectral Sparsification by Sign Descent},pdfauthor={Trevor Vaughn},pdfsubject={Partial colorings and spectral sparsification}}
\title{Linear-Size Spectral Sparsification by Sign Descent}
\author{Trevor Vaughn\thanks{Carnegie Mellon University. Email: tnvaughn@cmu.edu}}
\date{}
\begin{document}
\maketitle

\begin{abstract}
Given an undirected graph with $n$ vertices, $m$ edges, and arbitrary
positive weights, we construct a $(1\pm\varepsilon)$-spectral sparsifier
with $O(n/\varepsilon^2)$ edges. For every $0<\varepsilon<1$, the algorithm succeeds
with high probability and has worst-case running time
\[
 O\!\left(n+\varepsilon^{-3}\bigl[m\log^2 n+n\log^3 n\,(\log\log n)^4\bigr]\right).
\]
In each phase we seek a reweighting that brings many edge weights to
zero or twice their current value while keeping the relative change
in every graph energy small. Reversing the weight changes preserves
the error and exchanges deletion with doubling, so one choice deletes
at least half the edges that reach an endpoint. We balance a smooth
penalty for spectral error against a linear reward for alignment with
independent random signs. Each step changes the free weights by a
fixed fraction of their values at the start of the phase, with
directions chosen from the estimated signs of the negative gradient.
A weight that reaches an endpoint stays there for the rest of the phase.
We show that while few weights are fixed, the free edges admit a direction
whose gain in reward exceeds its first-order spectral cost by a constant.
The signs of the negative gradient achieve at least the same
first-order decrease. We bound the total decrease lost to sign errors.
A walk that froze too few coordinates would then drive the objective
below its lower bound.
We repeat the partial coloring on the surviving edges, choosing the
tolerances to keep the total relative error within $\varepsilon$.
\end{abstract}

\section{Introduction}\label{sec:introduction}

Let $G$ be a weighted undirected graph with Laplacian $L_G$. The
energy of an assignment $f$ of real values to the vertices is
\[
 f^\top L_G f=\sum_{uv\in E(G)} w_{uv}\bigl(f(u)-f(v)\bigr)^2.
\]
A spectral sparsifier uses a reweighted subset of the edges to
preserve this energy, up to small relative error, for every $f$.

\begin{theorem}\label{thm:main}
Let $G$ be an undirected graph with $n$ vertices, $m$ edges, and
positive real weights. For every $0<\eps<1$, a randomized algorithm
returns, with high probability, a reweighted subgraph $H$ with
$O(n/\eps^2)$ edges and
$(1-\eps)L_G\preceq L_H\preceq(1+\eps)L_G$.
Its worst-case running time is
\[
 O\!\left(n+\eps^{-3}\bigl[m\log^2 n+n\log^3 n\,(\log\log n)^4\bigr]\right).
\]
\end{theorem}

\paragraph{Removing edges in phases.}
A \emph{phase} computes and applies one partial coloring of the
current graph. Its goal is to delete a constant fraction of the
edges while keeping the relative change in every graph energy small.
If the phase starts with $p$ edges of weights $w_e$, write the new
weight of $e$ as $(1+x_e)w_e$ with $x_e\in[-1,1]$.
The energy change on an assignment $f$ is
$\sum_{e=uv}x_e w_e(f(u)-f(v))^2$. It is linear in $x$, so $x$ and
$-x$ have the same worst relative error. It therefore suffices to
bring a constant fraction of the coordinates to either endpoint of
$[-1,1]$. One of $x$ and $-x$ sets at least half the corresponding
weights to zero. We call such an $x$ a partial coloring.

The next phase starts from the new weights on the surviving edges.
Let $L$ be the Laplacian before the first coloring phase. We measure the errors
of all phases relative to $L$, and their sum bounds the final relative
error. A phase with tolerance $\rho$ may introduce relative error
$O(\rho)$. The partial coloring below allows $\rho=C\sqrt{n/p}$
for an absolute constant $C$.
Since $p$ decreases geometrically, the sum of these tolerances is
within a constant factor of the last one. Stopping at $O(n/\eps^2)$
edges therefore gives total error $O(\eps)$. In \Cref{sec:schedule}
we give some phases a larger share of the budget so that they can
take larger steps.

\paragraph{Choosing the weight changes.}
Let $\Delta L(x)$ be the change in the current Laplacian. On $\im L$,
define the normalized error matrix
\[
 H(x)=\frac{1}{\rho}\,L^{-1/2}\Delta L(x)L^{-1/2}.
\]
By the Rayleigh quotient, $\rho\|H(x)\|$ is the largest absolute
energy change relative to $L$. Thus it suffices to keep $\|H(x)\|$
bounded by a constant.

To compare possible moves, we need a smooth measure of spectral
error. The $\ell_\infty$-norm of $1$ and the eigenvalues of $H$ is
$\max\{1,\|H\|\}$. For an even $\ell=\Theta(\log n)$ with
$n^{1/\ell}\le2$, their $\ell$-norm $N(H)=(1+\Tr H^\ell)^{1/\ell}$
approximates this maximum within a constant factor. We use
$\log N(H)$ as the penalty, so the cost of a move depends on the
ratio of the new norm to the old one. The original weights
minimize this penalty, so we encourage movement toward the boundary
by rewarding alignment with a random sign vector $\sigma$. This gives
\[
 F(x)=\frac1\ell\log\bigl(1+\Tr H(x)^\ell\bigr)
       -\frac{\langle\sigma,x\rangle}{p}.
\]
To bound the change in the penalty, let $\lambda_i$ and $\lambda_i'$
be the eigenvalues of $H$ and $H+E$, respectively, in increasing
order. For every unit vector $v$ we have $|v^\top E v|\le\|E\|$, so
the min--max principle gives $|\lambda_i'-\lambda_i|\le\|E\|$ for
each $i$. By the triangle inequality for the $\ell$-norm,
\[
 |N(H+E)-N(H)|
 \le \Bigl(\sum_i|\lambda_i'-\lambda_i|^\ell\Bigr)^{1/\ell}
 \le n^{1/\ell}\|E\|\le2\|E\|.
\]
Both norms are at least one, and the derivative of $\log$ is at most
one on $[1,\infty)$. Thus adding $E$ to $H$ changes the penalty by
at most $2\|E\|$.

\paragraph{Why many weights reach an endpoint.}
Start at $x=0$. At each step we move the free coordinates a common
distance and freeze any coordinate that reaches $-1$ or $1$. The
signs of the negative gradient minimize the directional derivative of
$F$ over all directions supported on the free edges with entries in
$[-1,1]$. It therefore suffices to exhibit one such direction whose
gain in reward exceeds its marginal spectral cost.

Following every target sign could incur large spectral error. For
example, in the first phase either sign on a bridge changes the energy
of its cut by the full weight of the bridge, regardless of the other signs.
We show that a constant fraction of the target correlation is
attainable at small spectral cost.
With high probability over $\sigma$, for every sufficiently small
frozen set there is a direction $z\in\{-1,0,1\}^p$ supported on the
free coordinates with $\langle\sigma,z\rangle\ge p/2$ and relative
spectral error $O(\sqrt{n/p})$ (\Cref{lem:directions}).
We need the guarantee simultaneously for all small frozen sets
because the preceding steps depend on $\sigma$ and determine which
coordinates freeze.

Moving from $x$ to $x+tz$ for small $t>0$ gains at least $t/2$ in reward.
The error matrix changes by $tH(z)$, so the penalty increases by at
most $2t\|H(z)\|\le t/4$ when $\rho\ge C\sqrt{n/p}$ for a sufficiently
large $C$. Thus the directional derivative of $F$ along $z$ is at
most $-1/4$, and the negative gradient signs give at least as much
decrease. A sufficiently small step of length $h=\Theta(\rho^2/\log n)$
retains a constant fraction of this decrease. The objective starts
at zero and is bounded below by $-1$, since the penalty is
nonnegative and the reward is at most one. With exact signs the walk
must therefore freeze a constant fraction of the coordinates within
$O(1/h)$ steps. At that point $F\le0$ also bounds the penalty by
one, giving relative spectral error at most $e\rho$.

\paragraph{Estimating the edge derivatives.}
Let $E_e=\partial H/\partial x_e$. The update on edge $e$ compares
the reward derivative $\sigma_e/p$ with the marginal spectral cost
$\Tr(H^{\ell-1}E_e)/(1+\Tr H^\ell)$. All the numerators pair an edge
matrix with the same power of $H$. We estimate this power by a
random outer product $uv^\top$, whose pairing with an edge matrix
is $\Tr(uv^\top E_e)=v^\top E_e u$. To choose $u,v$, draw a vector
$\xi$ of independent uniform signs. For any
matrix $M$, $\E[\xi^\top M\xi]=\Tr M$, since the off-diagonal terms
have mean zero. Write $\ell=2k$, so that $H^{\ell-1}=H^{k-1}H^k$.
Compute $u=H^{k-1}\xi$ and take one more product by $H$ to obtain
$v=H^k\xi=Hu$. For every edge,
\[
 \E[v^\top E_e u]
 =\E[\xi^\top H^k E_e H^{k-1}\xi]
 =\Tr(H^k E_e H^{k-1})
 =\Tr(H^{\ell-1}E_e).
\]

To evaluate these bilinear forms on the graph, write
$H(x)=\Map\Delta L(x)\Map^\top/\rho$. Taking $\Map=L^{-1/2}$
gives the exact normalization. In \Cref{sec:graph} we approximate
this normalization with a Laplacian solver. Let $f=\Map^\top u$
and $g=\Map^\top v$ be the corresponding vertex functions.
Increasing $x_e$ by $t$ adds $tw_e$ to the
weight of $e=ab$. Applying the graph energy formula to this edge gives
\[
 v^\top E_e u
 =\frac{1}{\rho}\,g^\top\frac{\partial\Delta L(x)}{\partial x_e}\,f
 =\frac{w_e}{\rho}\bigl(f(a)-f(b)\bigr)\bigl(g(a)-g(b)\bigr).
\]
One scan of the edges now gives every numerator estimate from $f$
and $g$. The same probe estimates the common
denominator by $1+\|v\|_2^2$, since $\|v\|_2^2=\xi^\top H^\ell\xi$.
Consequently $(\sigma_e/p)(1+\|v\|_2^2)-v^\top E_e u$ has mean
$-(1+\Tr H^\ell)\partial_e F(x)$. A wrong sign on edge $e$
costs $2|\partial_e F(x)|$ in first-order decrease, so an error matters
in proportion to the magnitude of the derivative.

We estimate these signs over blocks of $B=\Theta(1/\rho)$ steps.
At the start of a block we average $B^2$ probes. At later steps we
estimate the change in the scaled derivatives over dyadic intervals.
For each correction we evaluate fresh random vectors at both endpoints,
pairing the two evaluations of each vector. On short intervals $H$
moves little, so the differences have small sampling error in expectation.
Reusing the corrections across steps gives all $B$ estimates in
$O(B^2)$ probe evaluations, the same order as the initial average
alone (\Cref{sec:steps}).

\paragraph{Comparison with prior work.}
Spielman and Srivastava~\cite{SS} sample edges independently with
probabilities proportional to their weights times their effective
resistances. This gives a nearly linear-time construction at the cost of
an extra logarithm in size, with $O(n\log n/\eps^2)$ edges.
Batson, Spielman, and Srivastava~\cite{BSS} first obtained
$O(n/\eps^2)$ edges deterministically by adding one edge at a time
between two spectral barriers. Lee and Sun~\cite{LS} attained this size
in nearly linear time by choosing batches of edges with a semidefinite
packing program. In their work bound the factor $\delta^{-3}$ for
solving this program already contributes six powers of $\log n$ at
their choice $\delta=\eps/\log^2 n$, compared with three in
\Cref{thm:main}~\cite[Theorem~4.5 and proof of Lemma~3.7]{LS}.

Lau, Wang, and Zhou~\cite{LWZ} gave a deterministic polynomial-time
discrepancy walk for the matrix partial-coloring theorem of Reis and
Rothvoss~\cite{RR}.
Jambulapati, Reis, and Tian~\cite{JRT} obtain the same edge count
by partial coloring. They use a bound on the Gaussian measure of an
operator-norm ball to show that projecting a Gaussian target onto its
intersection with the cube brings many coordinates to endpoints.
Here an elementary argument with random signs supplies the partial
coloring without using theorems from convex geometry. Up to iterated
logarithms, their running time is $O(m\eps^{-7/2}\log^4 m\log(1/\eps))$.
Our algorithm thus improves both the dependence on $1/\eps$ and the
logarithmic factors in the graph size.

\section{Preliminaries}\label{sec:preliminaries}

For symmetric matrices we write $\langle X,Y\rangle=\Tr(XY)$ and
$X\preceq Y$ when $Y-X$ is positive semidefinite (PSD). We use
$\|\cdot\|$ for the operator norm, $\|\cdot\|_F$ for the Frobenius
norm, and $|X|$ for the matrix obtained by taking absolute values of
the eigenvalues of a symmetric $X$. For $S\succeq0$, write
$S^\dagger$ for its pseudoinverse and $\Pi_S$ for the orthogonal projector
onto $\im S$.

\paragraph{Whitening.}
Suppose $S=\sum_i B_i$ is a sum of PSD matrices. Every $B_i$ vanishes
on $\ker S$, since for $f\in\ker S$ the nonnegative terms
$f^\top B_i f$ sum to zero. Multiplying each summand on both sides by
$S^{\dagger/2}=(S^\dagger)^{1/2}$ gives matrices
$A_i=S^{\dagger/2}B_iS^{\dagger/2}$ with $\sum_i A_i=\Pi_S$.
This change of coordinates is called \emph{whitening}. It turns
relative error into an operator norm bound. Indeed, for
$T=\sum_i x_i B_i$ we have
\[
 T=S^{1/2}\Bigl(\sum_i x_i A_i\Bigr)S^{1/2},\qquad
 -\eta S\preceq T\preceq\eta S
 \quad\Longleftrightarrow\quad
 \Bigl\|\sum_i x_i A_i\Bigr\|\le\eta .
\]

\paragraph{Edge differences.}
Orient each edge $e=ab$ arbitrarily and put $b_e=\mathbf e_a-\mathbf e_b$,
where $\mathbf e_a$ is the unit vector at $a$. The weighted incidence
matrix $C$ has rows $c_e^\top=\sqrt{w_e}\,b_e^\top$ and satisfies
$L=C^\top C$. Thus $(Cf)_e=\sqrt{w_e}(f(a)-f(b))$ for every vertex
function $f$.

\section{Finding a partial coloring}\label{sec:descent}

Let $A_1,\ldots,A_p$ be PSD matrices with $\sum_i A_i\preceq I$,
and write $\calA(x)=\sum_i x_i A_i$. A change $x$ in the coefficients
has matrix error $\calA(x)$. We seek $x\in[-1,1]^p$ with small
matrix error and a constant fraction of its coordinates at endpoints.
Such a vector is called a \emph{partial coloring}. We choose a tolerance $\rho$
for this call and require $\|\calA(x)\|=O(\rho)$.
A larger tolerance reduces the number of matrix products.

\begin{theorem}[Partial coloring]\label{thm:partial}
Suppose $A_1,\ldots,A_p\succeq0$ are $D\times D$ matrices with
$\sum_iA_i\preceq I$ and $\rank(\sum_iA_i)\le d$ for some $d\ge1$. Let
$64\sqrt{d/p}\le\rho\le1$. Choose an even integer $\ell\ge4$ with
$d^{1/\ell}\le2$. Except with probability $e^{-\ell}+e^{-p/64}$,
a randomized algorithm returns $y\in[-1,1]^p$ satisfying
\[
  \#\{i:y_i=-1\}\ge\gamma p/2,\qquad
  \|\calA(y)\|\le e^3\rho ,
\]
where $\gamma>0$ is an absolute constant. If a product by $\calA(x)$
costs $T_{\rm mv}\ge D$ and computing all $u^\top A_i v$ for a given
pair $u,v$ costs $T_{\rm sc}\ge p$, the time on every run is
\[
 O\!\left(\rho^{-3}\bigl(\ell^2 T_{\rm mv}+\ell T_{\rm sc}\bigr)\right).
\]
\end{theorem}

\paragraph{The objective.}
Choose independent uniform signs $\sigma_1,\ldots,\sigma_p$ and write
$\sigma=(\sigma_1,\ldots,\sigma_p)$. Keep these signs fixed throughout
the walk. For $H=\calA(x)/\rho$, put
\[
 Z(H)=1+\Tr H^\ell,\qquad
 \Phi(H)=\frac{\log Z(H)}{\ell},\qquad
 F(x)=\Phi(H)-\frac{\langle\sigma,x\rangle}{p}.
\]
The quantity $e^{\Phi(H)}$ is the $\ell$-norm of $1$ and the
nonzero eigenvalues of $H$, so it lies between $\max\{1,\|H\|\}$
and $(d+1)^{1/\ell}\max\{1,\|H\|\}$. The reward has absolute value
at most one on the cube. Thus $F(0)=0$, $F(x)\ge-1$, and
$\|\calA(x)\|=\rho\|H\|\le\rho e^{F(x)+1}$.

Let $S=\sum_i A_i$. If $v\in\ker S$, then
$0=v^\top S v=\sum_i\|A_i^{1/2}v\|_2^2$, so each $A_i v=0$.
Thus $\calA(x)$ vanishes on $\ker S$ for every $x$. We take traces
on $\im S$, of dimension at most $d$, with $X^0=I$ on this space.

To bound the derivatives of the penalty, we compare lower trace
powers with $\Tr X^\ell$. Let $\lambda_i$ be the eigenvalues of
a symmetric $X$ on this space. For $0<j<\ell$, H\"older's inequality gives
\begin{equation}\label{eq:moments}
 \Tr|X|^{\ell-j}
 =\sum_i (|\lambda_i|^\ell)^{1-j/\ell}
 \le d^{j/\ell}(\Tr X^\ell)^{1-j/\ell}
 \le 2^j Z(X).
\end{equation}
The factor $d^{j/\ell}$ is at most $2^j$ by the choice of $\ell$,
and $Z(X)=1+\Tr X^\ell$ bounds the remaining factor. The same
bound holds for $j=0$ since $\Tr X^\ell\le Z(X)$, and for $j=\ell$
since $\Tr I\le d\le2^\ell$.

To differentiate a matrix
power in a symmetric direction $E$, apply the product rule to
its $j$ factors,
\begin{equation}\label{eq:power-rule}
 \left.\frac{d}{ds}(X+sE)^j\right|_{s=0}
     =\sum_{a+b=j-1}X^a E X^b \qquad (j\ge1).
\end{equation}
For $j=\ell$, all $\ell$ terms have trace $\Tr(X^{\ell-1}E)$.
Differentiating $\Phi=\ell^{-1}\log Z$ therefore gives
\[
 D\Phi(X)[E]=\frac{\Tr(X^{\ell-1}E)}{Z(X)},\qquad
 |D\Phi(X)[E]|\le\frac{\Tr|X|^{\ell-1}}{Z(X)}\,\|E\|\le2\|E\|.
\]
The first inequality uses $|v^\top E v|\le\|E\|$ on each unit
eigenvector of $X$, and the second uses \eqref{eq:moments}.

\paragraph{A direction of decrease.}
While few coordinates have frozen, we need a direction on the
remaining coordinates whose reward gain exceeds its spectral cost.
The following lemma supplies such a direction for every sufficiently
small frozen set.

\begin{lemma}\label[lemma]{lem:directions}
Let $A_i\succeq0$, with $\sum_{i=1}^p A_i\preceq I$ of rank at
most $d$, and suppose $p\ge64d$. With probability at least
$1-e^{-p/64}$ over $\sigma$, for every $J\subseteq[p]$ of size at
most $\gamma p$ there is a vector $z\in\{-1,0,1\}^p$ satisfying
\[
 z_i=0\quad(i\in J),\qquad \langle\sigma,z\rangle\ge p/2,\qquad
 \|\calA(z)\|\le8\sqrt{d/p}.
\]
\end{lemma}

We prove the lemma in \Cref{sec:directions}. Suppose its conclusion
holds and $J$ is the current frozen set, with $|J|<\gamma p$.
Among directions in $[-1,1]^p$ vanishing on $J$, the negative
gradient signs maximize the first-order decrease, giving
$\sum_{i\notin J}|\partial_iF(x)|$. To lower bound this sum, use
the direction $z$ from the lemma. Its reward derivative is
$\langle\sigma,z\rangle/p$, and its penalty derivative is
$D\Phi(H)[\calA(z)/\rho]$. Thus
\begin{equation}\label{eq:slope}
 \sum_{i\notin J}|\partial_i F(x)|
 \ge -\langle\nabla F(x),z\rangle
 \ge \frac{\langle\sigma,z\rangle}{p}
       -\frac{2\|\calA(z)\|}{\rho}
 \ge \frac12-\frac{16}{\rho}\sqrt{\frac dp}\ge\frac14 .
\end{equation}
The algorithm only needs to estimate the signs of the derivatives.

\paragraph{Curvature of the penalty.}
To control the error in a first-order step, we bound the second
directional derivative of $\Phi$. For symmetric $X,E$ and integers
$a,b\ge0$, we use
\begin{equation}\label{eq:trace-sandwich}
 \bigl|\Tr(X^a E X^b E)\bigr|\le \|E\|^2\Tr|X|^{a+b}.
\end{equation}
In an eigenbasis of $X$, the trace is
$\sum_{i,j}\lambda_i^a\lambda_j^b E_{ij}^2$. Write
$s=a+b>0$. Weighted AM--GM and the symmetry $E_{ij}=E_{ji}$ give
\[
 \sum_{i,j}|\lambda_i|^a|\lambda_j|^b E_{ij}^2
 \le\sum_{i,j}\frac{a|\lambda_i|^s+b|\lambda_j|^s}{s}\,E_{ij}^2
 =\sum_{i,j}|\lambda_i|^s E_{ij}^2
 =\sum_i|\lambda_i|^s(E^2)_{ii}.
\]
Now $(E^2)_{ii}\le\|E\|^2$, which also proves the claim for
$a=b=0$.

Differentiate $D\Phi(X)[E]=\Tr(X^{\ell-1}E)/Z(X)$ in direction
$E$. The product rule \eqref{eq:power-rule} gives the numerator's
derivative, and $DZ(X)[E]=\ell\Tr(X^{\ell-1}E)$ gives the
denominator's derivative. Hence
\[
 D^2\Phi(X)[E,E]
 =\frac{\displaystyle\sum_{a+b=\ell-2}\Tr(X^a E X^b E)}{Z(X)}
   -\ell\left(\frac{\Tr(X^{\ell-1}E)}{Z(X)}\right)^2 .
\]
The last term is nonpositive. Each of the $\ell-1$ trace terms is at most
$\|E\|^2\Tr|X|^{\ell-2}\le4\|E\|^2 Z(X)$ by
\eqref{eq:trace-sandwich} and \eqref{eq:moments}. Thus, for all $X,E$,
\[
 D^2\Phi(X)[E,E]\le4(\ell-1)\|E\|^2.
\]

\paragraph{The walk.}
\Cref{alg:walk} gives the partial coloring. The target signs stay
fixed during a call, while \Cref{alg:scores} estimates the negative
gradient signs in blocks of $B$ steps. Since $1/h=\ell B^2$ is an
integer, the walk stays on a grid containing both endpoints of
$[-1,1]$. Freezing a coordinate as soon as it reaches an endpoint
therefore keeps the walk in the cube.

\begin{algorithm}[htbp]
\caption{\proc{PartialColor}$(A_1,\ldots,A_p;\rho,\ell)$}\label{alg:walk}
\begin{enumerate}[leftmargin=2.4em,itemsep=2pt,topsep=3pt]
\item Draw independent uniform signs $\sigma_i\in\{-1,1\}$;
set $x\gets0$ and $J\gets\varnothing$.
\item Let $B$ be the smallest power of two at least $C/\rho$,
for a sufficiently large absolute constant $C$. Set
$h\gets1/(\ell B^2)$ and $T\gets32/h$.
\item For $s=0,1,\ldots,T-1$, perform the following steps.
\begin{enumerate}[label=(\alph*),leftmargin=2em,itemsep=2pt,topsep=2pt]
\item Set $t\gets s\bmod B$; start a new estimation block if $t=0$.
\item Obtain $\widehat s_t$ from \Cref{alg:scores} at the current $x$.
\item Set $\Delta_i\gets0$ for $i\in J$ and
$\Delta_i\gets\sgn(\widehat s_{t,i})$ otherwise,
choosing either of $-1,1$ at a tie.
\item Set $x\gets x+h\Delta$ and $J\gets J\cup\{i:|x_i|=1\}$.
\item If $|J|\ge\gamma p$, return whichever of $x,-x$ has more
coordinates equal to $-1$.
\end{enumerate}
\item Report failure.
\end{enumerate}
\end{algorithm}

\paragraph{Progress in one step.}
At step $t$, let $J_t$ be the frozen set and $M_t$ the free
coordinates on which the direction $\Delta_t$ has the wrong sign.
Each wrong sign adds $2|\partial_iF(x_t)|$ to the directional
derivative. The slope bound \eqref{eq:slope} therefore gives
\[
 \langle\nabla F(x_t),\Delta_t\rangle
   =-\sum_{i\notin J_t}|\partial_i F(x_t)|+L_t
   \le -\tfrac14+L_t,\qquad
 L_t=2\sum_{i\in M_t}|\partial_i F(x_t)|.
\]
Set $L_t=0$ after the walk stops.

Since $\|\Delta_t\|_\infty\le1$ and $0\preceq\sum_i A_i\preceq I$,
we have $-\sum_i A_i\preceq\calA(\Delta_t)\preceq\sum_i A_i$.
Thus the matrix direction $E=\calA(\Delta_t)/\rho$ has norm at most
$1/\rho$. The reward is linear, so Taylor's theorem and the bound
on $D^2\Phi$ give
\begin{equation}\label{eq:progress}
 F(x_{t+1})-F(x_t)
 \le -h/4+hL_t+2(\ell-1)h^2/\rho^2
 \le -h/8+hL_t .
\end{equation}
The quadratic term is at most $2h/(\rho B)^2\le h/8$ by the
choice of $h$ and $C$.

\paragraph{Stopping the walk.}
Suppose $\sum_{t<T}L_t\le T/16$. Summing \eqref{eq:progress}
over any $j$ steps taken gives $F(x_j)\le hT/16-jh/8=2-jh/8$.
At $j=T$ this would give $F(x_T)\le-2$, contradicting $F\ge-1$.
Thus the walk reaches its freezing threshold before the step limit.

At stopping, $F(x)\le2$ gives
$\|\calA(x)\|\le\rho e^{F(x)+1}\le e^3\rho$.
The choice between $x$ and $-x$ preserves this bound and gives
at least $\gamma p/2$ entries equal to $-1$. We next bound the
total loss $\sum_{t<T}L_t$.

\subsection{Estimating the derivative signs}\label{sec:steps}

Fix the target $\sigma$ and write $X=\calA(x)/\rho$. Multiplying the
negative gradient by the positive scalar $\rho Z(X)$ gives
\[
 s_i(X)=-\rho Z(X)\partial_iF(x)
       =\frac{\rho\sigma_i}{p}Z(X)-\Tr(A_iX^{\ell-1}).
\]
We estimate all entries of $s(X)$ using the same random sign vector
$\xi\in\R^D$. With $k=\ell/2$, compute $u=X^{k-1}\xi$ and
$v=X^k\xi=Xu$. The probe is
\begin{equation}\label{eq:probe}
 P_i(X,\xi)=\frac{\rho\sigma_i}{p}\bigl(1+\|v\|_2^2\bigr)
                          -u^\top A_i v .
\end{equation}
Indeed, $\E[\xi^\top M\xi]=\Tr M$ gives $\E\|v\|_2^2=\Tr X^\ell$
and $\E[u^\top A_i v]=\Tr(X^{k-1}A_iX^k)=\Tr(A_iX^{\ell-1})$.
Thus $\E P(X,\xi)=s(X)$. A probe uses $k$ matrix products and
one evaluation of all the bilinear forms.

\paragraph{Maintaining the estimates.}
A wrong sign in an estimate $\widehat s_i$ entails
$|\widehat s_i-s_i(X)|\ge|s_i(X)|$. Summing over the wrong signs
relates the error in $\widehat s$ directly to the loss in the walk,
\begin{equation}\label{eq:sign-loss}
 L_t=2\sum_{i\in M_t}\frac{|s_i(X)|}{\rho Z(X)}
      \le\frac{2\|\widehat s-s(X)\|_1}{\rho Z(X)} .
\end{equation}
Divide the walk into blocks of $B$ steps and let $H_t$ be the
matrix just before step $t$ of a block, $0\le t<B$. The value
$s(H_t)$ equals $s(H_0)$ plus the changes in $s$ over any
partition of the first $t$ steps. Short intervals are cheap to
estimate because the same random vector gives similar probes at
their two endpoints. The estimation errors add along the
partition, so we choose it to contain at most one interval of
each dyadic length.

\Cref{alg:scores} maintains this partition. At $t=0$ it estimates
$s(H_0)$ directly. For $t>0$, subtracting the largest power of two
$b$ dividing $t$ clears its lowest nonzero bit. The old estimate at $t-b$ already
covers the prefix through that step. The coefficient vector stored
with it lets us evaluate fresh probes at $H_{t-b}$ and pair them with
probes at $H_t$. Both states are fixed before the draw, so the
correction has conditional mean $s(H_t)-s(H_{t-b})$. The interval
$[t-b,t]$ replaces any shorter intervals after $t-b$ in the partition.

\begin{algorithm}[htbp]
\caption{Estimating the scores at step $t$ of a block}\label{alg:scores}
\begin{enumerate}[leftmargin=2.4em,itemsep=2pt,topsep=3pt]
\item Use the current $x$ to represent $H_t=\calA(x)/\rho$ implicitly.
\item If $t=0$, clear the previous block's records and draw $B^2$
fresh independent uniform sign vectors
$\xi_j\in\{-1,1\}^D$. Using the probe in \eqref{eq:probe}, set
$\widehat s_0\gets B^{-2}\sum_{j=1}^{B^2}P(H_0,\xi_j)$.
\item If $t>0$, perform the following steps.
\begin{enumerate}[label=(\alph*),leftmargin=2em,itemsep=2pt,topsep=2pt]
\item Let $b$ be the largest power of two dividing $t$ and
set $q_b\gets\lceil B^{1/2}b^{3/2}\rceil$.
\item Draw $q_b$ fresh independent uniform sign vectors $\xi_j$;
use the same vector at both endpoints of each difference.
\item Set $\widehat s_t\gets\widehat s_{t-b}
+\frac1{q_b}\sum_{j=1}^{q_b}
\bigl(P(H_t,\xi_j)-P(H_{t-b},\xi_j)\bigr)$.
\item Discard stored records with indices greater than $t-b$.
\end{enumerate}
\item Store $(x,\widehat s_t)$ under index $t$ and return $\widehat s_t$.
\end{enumerate}
\end{algorithm}

\paragraph{How much a probe changes.}
Fix the history at the start of a block and put $Z_0=Z(H_0)$.
To use \eqref{eq:sign-loss}, first compare $Z(H_t)$ with $Z_0$.
Each step moves $H$ by at most $h/\rho$, so all its states lie in
the ball $\|X-H_0\|\le Bh/\rho$ on the image of $\sum_i A_i$.
For any $X$ in this ball, integrate $|D\Phi(W)[E]|\le2\|E\|$ along
the segment from $H_0$ to $X$. Since $\log Z=\ell\Phi$,
\begin{equation}\label{eq:local-z}
 |\log Z(X)-\log Z_0|\le 2\ell Bh/\rho=2/(\rho B)\le\log 2 .
\end{equation}
Thus $Z_0/2\le Z(X)\le2Z_0$. By \eqref{eq:sign-loss}, expected
$\ell_1$ error $O(Z_0/B)$ then gives expected loss
$O(1/(\rho B))=O(1/C)$ per step.

For $q$ fresh sign vectors, write
$e_q(X)=q^{-1}\sum_{j=1}^q P(X,\xi_j)-s(X)$ for the error of
the probe average at $X$. To estimate the change $s(Y)-s(X)$,
average $P(Y,\xi_j)-P(X,\xi_j)$ using the same vectors at both
matrices. Subtracting the true change gives the correction error
\[
 \frac1q\sum_{j=1}^q\bigl(P(Y,\xi_j)-P(X,\xi_j)\bigr)
 -\bigl(s(Y)-s(X)\bigr)=e_q(Y)-e_q(X).
\]

\begin{lemma}\label[lemma]{lem:probe-variation}
Let $X,Y$ be fixed symmetric matrices on $\im(\sum_i A_i)$,
extended by zero on its orthogonal complement, with
$\|X-H_0\|,\|Y-H_0\|\le Bh/\rho$. For $q$ independent sign vectors,
\begin{align}
 \E\|e_q(X)\|_1 &= O\bigl(Z_0/\sqrt q\bigr),\label{eq:probe-size}\\
 \E\|e_q(Y)-e_q(X)\|_1
      &= O\bigl(Z_0\,\ell\|Y-X\|/\sqrt q\bigr).\label{eq:probe-change}
\end{align}
\end{lemma}
We prove \Cref{lem:probe-variation} at the end of the subsection.

\paragraph{Accuracy and work in one block.}
The $B^2$ probes at $H_0$ give expected error $O(Z_0/B)$ by
\eqref{eq:probe-size}. For a correction spanning $b$ steps, put
$r=b/B$. Each of the $b$ steps changes $H$ by at most $h/\rho$,
so the factor $\ell\|Y-X\|$ in \eqref{eq:probe-change} satisfies
\[
 \ell\|H_t-H_{t-b}\|\le \ell b h/\rho=r/(\rho B)\le r .
\]
Thus $q$ probe pairs give expected error $O(Z_0r/\sqrt q)$,
conditional on the history just before the draw. This bound also
holds conditional on the start of the block, since its right-hand
side depends only on $Z_0$ and the interval length.

A single estimate uses at most one correction of length $b$.
A full block computes $B/(2b)=1/(2r)$ such corrections, each
using $2q_b$ probe evaluations. To see how many probes to use, try
$q_b$ of order $B^2r^\nu$. One correction then contributes expected
error $O((Z_0/B)r^{1-\nu/2})$, while all corrections of this length
cost $O(B^2r^{\nu-1})$ evaluations. Both bounds decay geometrically
with the scale for $1<\nu<2$. Taking $\nu=3/2$ gives
\[
 \begin{aligned}
 \text{expected error in one correction}
     &=O\left(\frac{Z_0r}{\sqrt{q_b}}\right)
       =O\bigl((Z_0/B)r^{1/4}\bigr),\\
 \text{evaluations for this length}
     &=q_b/r=O(B^2r^{1/2}).
 \end{aligned}
\]
Both bounds decrease geometrically as $r$ runs through
$1/2,1/4,\ldots,1/B$. Rounding $q_b$ up changes the work by at
most a factor of two, since $B^2r^{3/2}\ge1$.

For a fixed $t$, let $I_t$ be the set of right endpoints of the
intervals in its partition, and let $\delta_u$ be the sampled
correction ending at $u$ minus its conditional mean. If that correction spans an interval
$[u-b,u]$, its conditional mean is $s(H_u)-s(H_{u-b})$. Adding
these means over the partition telescopes to $s(H_t)-s(H_0)$.
Thus the recurrence in \Cref{alg:scores} expresses $\widehat s_t$ as the initial
estimate plus this true change and the sum of the correction errors.
Subtracting $s(H_t)$ gives
\[
 \widehat s_t-s(H_t)=e_{B^2}(H_0)+\sum_{u\in I_t}\delta_u .
\]
Here $I_t$ depends only on $t$, even though the states at its
endpoints depend on previous draws. Taking norms and using the
conditional error bound once for each length in the partition gives
\begin{equation}\label{eq:estimate-error}
 \E\bigl[\|\widehat s_t-s(H_t)\|_1\mid\text{start of block}\bigr]
 \le \frac{O(Z_0)}{B}
      \left(1+\sum_{a=1}^{\log_2 B}2^{-a/4}\right)=O(Z_0/B).
\end{equation}
For this calculation, a block that stops early can be completed at
its last state with zero loss on the remaining steps. Summing the
work bound over $r=1/2,1/4,\ldots,1/B$ gives $O(B^2)$ evaluations
including the initial average. The stored indices are $0$ and the
right endpoints in $I_t$. Each record holds two $p$-vectors, so the
space beyond that for a probe is $O(p\log B)$.

\paragraph{Total loss over the walk.}
By \eqref{eq:sign-loss}, \eqref{eq:local-z}, and \eqref{eq:estimate-error},
\[
 \E[L_t\mid\text{start of block}]
 \le\frac{4}{\rho Z_0}
      \E[\|\widehat s_t-s(H_t)\|_1\mid\text{start of block}]
 =O(1/(\rho B))\le 1/64
\]
for sufficiently large $C$. The estimates within a block share
probes, so we apply concentration to the average loss $U_j$ in
each block $j$. Given the preceding blocks, $\E U_j\le1/64$.
Also $0\le U_j\le6/\rho$ on every run. Indeed, in any direction
$\|\Delta\|_\infty\le1$ the penalty derivative has magnitude at
most $2/\rho$ and the reward derivative at most one, giving
$L_t\le2\|\nabla F\|_1\le2(2/\rho+1)\le6/\rho$.

There are $N=T/B=32\ell B$ blocks, counting zeros after stopping.
Since $\sum_{t<T}L_t=B\sum_jU_j$, the stopping argument after
\eqref{eq:progress} requires $\sum_jU_j\le N/16$. Since $e^y\le1+2y$ on $[0,1]$,
\[
 \E[e^{\rho U_j/6}\mid\text{past blocks}]
 \le 1+\tfrac\rho3\E[U_j\mid\text{past blocks}]\le e^{\rho/192}.
\]
Successive conditioning gives
$\E\exp((\rho/6)\sum_j U_j)\le e^{N\rho/192}$. Markov's
inequality at $\sum_j U_j>N/16$ therefore gives
\[
 \Prob\!\left\{\sum_{t<T}L_t>T/16\right\}
 \le e^{-N\rho/96}e^{N\rho/192}
 =e^{-\ell\rho B/6}\le e^{-\ell}.
\]
This holds for every fixed $\sigma$. Together with
\Cref{lem:directions} and the stopping argument, it gives the
partial coloring of \Cref{thm:partial} with failure probability
at most $e^{-\ell}+e^{-p/64}$. The $N$ blocks use
$O(NB^2)=O(\ell\rho^{-3})$ probes, each costing
$O(\ell T_{\rm mv}+T_{\rm sc})$.

\begin{proof}[Proof of \Cref{lem:probe-variation}]
Write $c_i=\rho\sigma_i/p$. At a matrix $W$ put $U=W^{k-1}$ and
$V=W^k$. Then \eqref{eq:probe} reads
\[
 P_i(W,\xi)=c_i+\xi^\top M_i(W)\xi,\qquad
 M_i(W)=c_iV^\top V-U^\top A_iV .
\]
At $X$ the estimation error comes from $\xi^\top M_i(X)\xi$.
In a paired correction the constants cancel, and the corresponding
quadratic form is $\xi^\top(M_i(Y)-M_i(X))\xi$.

For a matrix $M$, expand $\xi^\top M\xi$ into diagonal and
off-diagonal terms. Since $\xi_a^2=1$, the diagonal terms sum to
$\Tr M$. Pairing the two terms for each $a<b$ therefore gives
\[
 \xi^\top M\xi-\Tr M
 =\sum_{a<b}(M_{ab}+M_{ba})\xi_a\xi_b .
\]
Terms from distinct pairs are uncorrelated: their product contains
an independent sign appearing only once, and so has expectation zero.
Thus the variance is
$\sum_{a<b}(M_{ab}+M_{ba})^2\le2\|M\|_F^2$. Averaging $q$
independent copies divides the variance by $q$. By Cauchy--Schwarz,
\begin{equation}\label{eq:sample-mean}
 \E\left|\frac1q\sum_{j=1}^q\xi_j^\top M\xi_j-\Tr M\right|
 \le\sqrt{\frac{\operatorname{Var}(\xi^\top M\xi)}q}
 \le\sqrt{2/q}\,\|M\|_F .
\end{equation}

Summing \eqref{eq:sample-mean} over $i$, it remains to bound
$\sum_i\|M_i(X)\|_F$ and $\sum_i\|M_i(Y)-M_i(X)\|_F$.

For any matrices $U,V$ of compatible dimensions,
\begin{equation}\label{eq:bilinear}
 \sum_i\|U^\top A_i V\|_F
 \le \left(\sum_i\|A_i^{1/2}U\|_F^2\right)^{1/2}
     \left(\sum_i\|A_i^{1/2}V\|_F^2\right)^{1/2}
 \le \|U\|_F\|V\|_F .
\end{equation}
For the first inequality, factor $A_i$ to obtain
\[
 \|U^\top A_i V\|_F
 =\|(A_i^{1/2}U)^\top(A_i^{1/2}V)\|_F
 \le\|A_i^{1/2}U\|_F\|A_i^{1/2}V\|_F .
\]
Summing over $i$ and applying Cauchy--Schwarz gives the stated bound.
For the second, $\sum_i\|A_i^{1/2}U\|_F^2
=\Tr(U^\top(\sum_iA_i)U)\le\|U\|_F^2$, and likewise for $V$.
For the other term in $M_i(W)=c_iV^\top V-U^\top A_iV$,
the coefficients satisfy $\sum_i|c_i|=\rho\le1$. Thus
\[
 \sum_i\|c_iV^\top V\|_F
 =\rho\|V^\top V\|_F
 \le\rho\Tr(V^\top V)
 =\rho\|V\|_F^2 .
\]
Together with \eqref{eq:bilinear}, this reduces the bound on
$\sum_i\|M_i(W)\|_F$ to bounds on the two powers $U,V$.

For $j=k-1,k$ and $W=X+s(Y-X)$, $0\le s\le1$, the segment lies
in the same ball and \eqref{eq:moments} gives
$\|W^j\|_F^2=\Tr|W|^{2j}=O(Z_0)$ because
$2j\in\{\ell-2,\ell\}$. In particular
$\sum_i\|M_i(X)\|_F\le \rho\|V\|_F^2+\|U\|_F\|V\|_F=O(Z_0)$
at $W=X$, proving \eqref{eq:probe-size}.

For the difference, differentiate along the segment and put
$E=Y-X$. Write $\dot U=dU/ds$ and $\dot V=dV/ds$.
By \eqref{eq:power-rule}, the derivative of $W^j$ is
a sum of $j$ terms $W^a E W^b$ with $a+b=j-1$. Each satisfies
\[
 \|W^a E W^b\|_F^2
 =\Tr(W^b E W^{2a}E W^b)
 =\Tr(W^{2a}E W^{2b}E)
 \le \|E\|^2\Tr|W|^{2j-2}
 =O(\|E\|^2 Z_0)
\]
by \eqref{eq:trace-sandwich} and \eqref{eq:moments}, since
$2j-2\in\{\ell-4,\ell-2\}$. Thus $\dot U$ and $\dot V$ both
have Frobenius norm $O(\ell\|E\|\sqrt{Z_0})$. The product rule gives
\[
 \frac{d}{ds}M_i(W)
 =c_i(\dot V^\top V+V^\top\dot V)
    -\dot U^\top A_iV-U^\top A_i\dot V .
\]
Using \eqref{eq:bilinear} and $\sum_i|c_i|=\rho$, we obtain
\[
 \begin{aligned}
 \sum_i\left\|\frac{d}{ds}M_i(W)\right\|_F
 &\le 2\rho\|\dot V\|_F\|V\|_F
       +\|\dot U\|_F\|V\|_F+\|U\|_F\|\dot V\|_F\\
 &=O(\ell\|E\|Z_0).
 \end{aligned}
\]
Integration from $0$ to $1$ gives the same bound for
$\sum_i\|M_i(Y)-M_i(X)\|_F$. Applying
\eqref{eq:sample-mean} proves \eqref{eq:probe-change}.
\end{proof}

\subsection{A signed direction on the free coordinates}\label{sec:directions}

The frozen set in \Cref{lem:directions} may depend on $\sigma$.
Fix for now a set $J$ with $|J|\le p/8$. We construct $z$ by
accepting or skipping each sign outside $J$, setting $z_i=\sigma_i$
on acceptance and $z_i=0$ everywhere else. Its correlation
$\langle\sigma,z\rangle$ counts the accepted signs. We aim for
$3p/4$ acceptances in expectation while satisfying the norm bound
on every outcome. The excess above $p/2$ will let us make the
correlation bound simultaneous over all small $J$.

\paragraph{Keeping the signed sum between two bounds.}
Put $\tau=4\sqrt{d/p}\le1/2$. Just before index $i$, let
$M=\sum_{j<i}z_jA_j$ be the signed sum and $S=\sum_{j<i}A_j$ the
sum of all matrices inspected so far. We keep $M$ between
$-\tau(I+S)$ and $\tau(I+S)$, working on $\im(\sum_i A_i)$.
This allows a final norm of at most $2\tau=8\sqrt{d/p}$, since
$\sum_i A_i\preceq I$. Including $A_i$ in $S$ increases the
slack on each side by $\tau A_i$, even when we skip its sign.

Let $X_+=\tau(I+S)-M$ and $X_-=\tau(I+S)+M$ be the two slacks
before index $i$. Both start at $\tau I$. For either slack $X$,
the size of $A_i$ relative to $X$ is
$\|X^{-1/2}A_iX^{-1/2}\|\le\Tr(A_iX^{-1})$. Before exposing
$\sigma_i$, accept $i$ if $i\notin J$ and
\[
 a_i=\Tr\bigl(A_i(X_+^{-1}+X_-^{-1})\bigr)\le\tau .
\]
On acceptance set $z_i=\sigma_i$, and otherwise set $z_i=0$.
We will show that $\E X_\pm^{-1}\preceq\tau^{-1}I$ throughout
the construction. This gives $\E a_i\le(2/\tau)\Tr A_i$.
Each rejection outside $J$ requires $a_i>\tau$, so Markov's
inequality gives
\[
 \E\,\#\{i\notin J:z_i=0\}
 \le \frac1{\tau}\sum_i \E a_i
 \le \frac{2}{\tau^2}\Tr\sum_i A_i
 \le \frac{2d}{\tau^2}=\frac p8 .
\]

We verify the inverse bound together with positivity of the slacks. On rejection the extra
slack $\tau A_i$ can only decrease either inverse. On acceptance,
fix one of the slacks $X$ and write $Q=X^{-1/2}A_iX^{-1/2}$.
After including the target sign, the new slack is
$X^{1/2}(I+\tau Q\pm Q)X^{1/2}$ with either sign equally likely.
Since $Q$ is PSD, each eigenvalue $b$ satisfies
$0\le b\le\Tr Q\le a_i\le\tau\le1/2$. On an eigenvector of $Q$
with eigenvalue $b$, the two possible matrices $I+\tau Q\pm Q$
have eigenvalues $1+\tau b+b$ and $1+\tau b-b$, both positive.
Their inverses therefore have the reciprocal eigenvalues, and the
average of the two inverses has eigenvalue
\[
 \frac12\left(\frac1{1+\tau b+b}+\frac1{1+\tau b-b}\right)
 =\frac{1+\tau b}{(1+\tau b)^2-b^2}
 =\left(1+\tau b-\frac{b^2}{1+\tau b}\right)^{-1}\le1 .
\]
Here $\tau b\ge b^2/(1+\tau b)$ because $b\le\tau$. Conjugating
by $X^{-1/2}$ gives $\E[X_{\rm new}^{-1}\mid\text{past}]\preceq X^{-1}$.
Induction proves positivity and $\E X_\pm^{-1}\preceq\tau^{-1}I$
at every index. The count above now applies. At most $p/8$ indices
are excluded by $J$ and at most $p/8$ others are skipped in
expectation. The expected correlation is therefore at least $3p/4$.

\paragraph{A direction for every small frozen set.}
For fixed $J$, let $W_J(\sigma)$ be the maximum of $\langle\sigma,z\rangle$
over all $z\in\{-1,0,1\}^p$ vanishing on $J$ with
$\|\calA(z)\|\le2\tau$.
The construction gives $\E W_J\ge3p/4$. The feasible set is
independent of $\sigma$, and flipping one sign changes every
candidate's correlation by at most two. The maximum changes by at
most two as well. McDiarmid's inequality,
applied to the $p$ independent signs with sensitivity $2$ for each,
therefore bounds a downward deviation of $p/4$ by
\[
 \Prob\{W_J<p/2\}
 \le\Prob\{W_J-\E W_J\le-p/4\}
 \le\exp\!\left(-\frac{2(p/4)^2}{\sum_{i=1}^p 2^2}\right)
 =e^{-p/32}.
\]
Choose $\gamma\le1/8$ so that $\gamma\log(e/\gamma)\le1/64$.
The number of eligible frozen sets is at most
\[
 \sum_{j\le\gamma p}\binom pj
 \le\gamma^{-\gamma p}\sum_{j=0}^p\binom pj\gamma^j
 =\gamma^{-\gamma p}(1+\gamma)^p
 \le(e/\gamma)^{\gamma p}\le e^{p/64}.
\]
A union bound shows that every such $J$ has $W_J\ge p/2$
with probability at least
$1-e^{p/64}e^{-p/32}=1-e^{-p/64}$.

\section{Iterating the partial coloring}\label{sec:schedule}

One partial coloring removes a constant fraction of the summands
while introducing a bounded matrix error. We repeat it on the
surviving summands until only $O(d/\eps^2)$ remain. To obtain a
relative error of at most $\eps$, we measure every phase's change
against the same initial sum and keep their total within an error
budget. Each phase also needs its input sum bounded by $I$, so the
budget must preserve that condition throughout. Finally, larger
phase tolerances make the colorings faster; we allocate the available
error to control both the cost of scanning summands and the cost of
the solver used in the graph algorithm.

Let $S_0=\sum_{i=1}^{m_0}A_i$, where $A_i\succeq0$, $\rank S_0\le d$,
and the nonzero eigenvalues of $S_0$ lie in $[\alpha,1/2]$ for an
absolute constant $0<\alpha\le1/2$. We will reweight these same
summands in every phase. Keeping $\|S-S_0\|\le\alpha\eps$ for
every intermediate sum $S$ suffices for two reasons. First, $S$
vanishes on $\ker S_0$ and $S_0\succeq\alpha\Pi_{S_0}$, so for all $f$,
\[
 |f^\top(S-S_0)f|
 =|(\Pi_{S_0}f)^\top(S-S_0)(\Pi_{S_0}f)|
 \le\alpha\eps\|\Pi_{S_0}f\|_2^2\le\eps f^\top S_0f .
\]
Second, $\|S\|\le1/2+\alpha\eps\le1$, as required for the next
application of \Cref{thm:partial}.

Start with weights $w_i=1$. In phase $t$ let $S_t=\sum_i w_i A_i$
and let $p_t$ count the positive weights. Stop when $p_t\le C_0d/\eps^2$.
Otherwise apply \Cref{thm:partial} to the positive-weight summands
$w_i A_i$ with tolerance $\rho_t$ chosen below, reporting failure
if the call does. Given a coloring $y$, multiply the corresponding weights by
$1+y_i$. At least $\gamma p_t/2$ of these multipliers vanish.
Set $q=1-\gamma/2$. Since
$S_{t+1}-S_t=\sum_{i:w_i>0}y_iw_iA_i$, we obtain
\begin{equation}\label{eq:phase}
 p_{t+1}\le q p_t,\qquad \|S_{t+1}-S_t\|\le e^3\rho_t.
\end{equation}
Every returned coloring gives the first inequality by the stopping
rule. The second follows from its spectral guarantee on success.
If $m_0>C_0d/\eps^2$, let $K$ be the least integer
such that $m_0q^K\le C_0d/\eps^2$. Since $p_t\le m_0q^t$, we
make at most $K$ calls.

\paragraph{Choosing the tolerances.}
To keep $\|S_{j+1}-S_0\|\le\alpha\eps$, it suffices by
\eqref{eq:phase} that $\sum_{t=0}^j\rho_t\le\alpha\eps/e^3$.
Each call to \Cref{thm:partial} requires $\rho_t\ge64\sqrt{d/p_t}$.
We first check that these minimum tolerances fit within the total
error budget, leaving room to increase them later. Each earlier term
is at most a geometrically decaying multiple of the last one, so their
sum up to any active phase $j$ is controlled by that last term. Indeed,
$p_j\le q^{j-t}p_t$ and $p_j>C_0d/\eps^2$ give
\[
 \sum_{t=0}^j\sqrt{d/p_t}
 \le\sqrt{d/p_j}\sum_{t=0}^j q^{(j-t)/2}
 \le\frac{\sqrt{d/p_j}}{1-\sqrt q}
 \le\frac{\eps}{\sqrt{C_0}(1-\sqrt q)} .
\]
Choose $C_0$ large enough that the required tolerances use at most
half the available error.

We spend the remaining budget on the two costs of a matrix product:
scanning the $p_t$ surviving summands and applying the fixed solver.
For the scans, use an allowance proportional to $\eps q^{t/4}$.
Since $p_t\le m_0q^t$ and $q^t/(q^{t/4})^3=q^{t/4}$, both the
added tolerances and the resulting work bounds sum geometrically.
For the solver cost, divide its allowance equally among the at
most $K$ phases. Thus take
\begin{equation}\label{eq:schedule}
 \rho_t=64\sqrt{d/p_t}
       +\frac{\eps}{C_1}\left(q^{t/4}+\frac1K\right).
\end{equation}
The two allowances sum to at most
$(\eps/C_1)((1-q^{1/4})^{-1}+1)$. Taking $C_1$ large enough
therefore gives $\sum_{t=0}^j\rho_t\le\alpha\eps/e^3$ for every
prefix, and in particular $\rho_t\le1$. Induction using
\eqref{eq:phase} now gives $\|S_{j+1}-S_0\|\le\alpha\eps$ after
every successful phase. This justifies each call and proves the
final relative error bound.

The two allowances in \eqref{eq:schedule} give the respective work bounds
\begin{equation}\label{eq:sum-work}
 \begin{aligned}
 \sum_t p_t\rho_t^{-3}
 &\le (C_1/\eps)^3 m_0 \sum_{t=0}^{K-1}q^{t/4}=O(m_0\eps^{-3}),\\
 \sum_t \rho_t^{-3}
 &\le K(C_1K/\eps)^3=O(K^4\eps^{-3}).
 \end{aligned}
\end{equation}

\section{Sparsifying a graph}\label{sec:graph}

We apply \Cref{sec:schedule} with $d=n$ and accuracy $\eps/4$, so
the stopping threshold is $16C_0 n/\eps^2$ edges. Discard loops
and combine each parallel class onto one representative in
$O(n+m)$ time. Return the graph if it meets the threshold.

\paragraph{Sampling a Laplacian.}
We use the scalar case of the sampling routine of Kyng et
al.~\cite[full version, Lemma~H.1]{KLPSS}. For an input Laplacian $L_*$,
add a diagonal shift $\zeta I$ with $\zeta=w_{\min}/(16n^2)$, where
$w_{\min}$ is its smallest edge weight. Bounding squared differences
by the energy along paths of length at most $n-1$ within each
component gives a lower bound of $w_{\min}/n^2$ on the smallest
positive eigenvalue of $L_*$.

Run the sampler with a constant reduction factor and accuracy
$\delta/8$ for a desired $0<\delta\le1$. The sampler retains the
shift and reweights a subset of the input edges. For the resulting
Laplacian $\widehat L$, its guarantee gives, on $\im L_*$,
\[
 -(e^{\delta/8}-1)(L_*+\zeta I)
 \preceq \widehat L-L_*
 \preceq (e^{\delta/8}-1)(L_*+\zeta I).
\]
Since $L_*+\zeta I\preceq2L_*$ on this space, the relative error is
at most $2(e^{\delta/8}-1)\le\delta/2\le 1-e^{-\delta}$.
Both Laplacians vanish on $\ker L_*$, so
$e^{-\delta}L_*\preceq\widehat L\preceq e^\delta L_*$.
The sample has $O(n\log n/\delta^2)$ edges with high probability.

\paragraph{Reducing the number of phases.}
To start within a factor $O(\log n)$ of the desired edge count,
first sample the input if it has more than $n\log n/\eps^2$ edges.
Use the sampler above with $\delta=\eps/8$. With high probability
the sampled Laplacian $L$ satisfies
\[
 e^{-\eps/8}L_G\preceq L\preceq e^{\eps/8}L_G .
\]
If sampling is unnecessary, take $L=L_G$. In either case $L$ uses
$m_0=O(n\log n/\eps^2)$ edges of the input. Return this graph if
it meets the threshold. Otherwise \eqref{eq:phase} brings the
edge count below the threshold within $K=O(\log\log n)$ phases.

\paragraph{Normalizing the edge matrices.}
We seek a map $\Map$ such that $S_0=\Map L\Map^\top$ has its
positive eigenvalues between a constant and $1/2$, as required by
\Cref{sec:schedule}. For the weighted incidence matrix $C_L$ of
$L$, consider first the exact map $R_0=C_LL^\dagger$. Since
$C_L^\top C_L=L$, we have
\[
 R_0^\top R_0
 =L^\dagger C_L^\top C_LL^\dagger
 =L^\dagger LL^\dagger
 =L^\dagger,
 \qquad
 R_0 L R_0^\top=C_LL^\dagger C_L^\top=\Pi_{\im C_L} .
\]
The first identity says that $R_0$ has the same effect on vector
norms as $L^{\dagger/2}$, while the second says that its normalized
edge matrices sum to a projector. This representation lets us use
an approximate inverse in place of an inverse square root. Each
application of $R_0$, however, requires a product by $L^\dagger$ and a scan of the edges
of $L$. We instead use a Laplacian solver and the incidence matrix of a
sparse graph to construct a map that costs $O(n\log n)$ to apply.
Its normalized sum has positive eigenvalues bounded above and
below by constants.

Apply the multigrid construction of Kyng et
al.~\cite[full version, Lemmas~I.3 and~I.5]{KLPSS} to $L+\zeta I$
with the shift above and a sufficiently large constant recursion
parameter. With an exact solve on the final constant-size system,
their Lemma~3.3 and the error bounds in the proof of Lemma~I.3 give
a symmetric operator that
approximates $(L+\zeta I)^{-1}$ within a factor of
$\exp(4\sum_{i\ge0}(i+8)^{-2})<e$. Project onto $\im L$ before
and after each application by subtracting component means. Since
$L+\zeta I\preceq(17/16)L$ on $\im L$, the projected operator
$W$ satisfies the first bound below.

For the second bound, take $D$ to be the weighted incidence matrix
of the graph produced by the first sparsification in the solver
construction, with the shift removed. This step uses accuracy
$1/64$ on the shifted system, which corresponds to $\delta=1/8$
in the sampling bound above. If the
construction skips this step because the input is already sparse,
take $D=C_L$. With high probability $D$ has $O(n\log n)$ rows and
\begin{equation}\label{eq:solver}
 \tfrac13 L^\dagger\preceq W\preceq3L^\dagger,\qquad
 \tfrac12 L\preceq D^\top D\preceq2L .
\end{equation}
On a successful construction, applying $W$ takes $O(n\log n)$
operations. We report failure if an
application exceeds $C_{\rm sol}n\log n$ operations, for a sufficiently
large absolute constant $C_{\rm sol}$.

Set $\Map=DW/8$ and $Q=L^{1/2}WL^{1/2}$. For $f\in\im L$,
the bounds on $D^\top D$ give
\[
 \frac{\|Qf\|_2^2}{128}
 \le\|\Map L^{1/2}f\|_2^2
 =\frac1{64}f^\top L^{1/2}WD^\top DWL^{1/2}f
 \le\frac{\|Qf\|_2^2}{32}.
\]
The bounds on $W$ give $\tfrac13 I\preceq Q\preceq3I$ on
$\im L$. Thus the squared singular values of $\Map L^{1/2}$
on this space lie in $[1/1152,9/32]$. These are the positive
eigenvalues of $S_0=\Map L\Map^\top$, so the conditions of
\Cref{sec:schedule} hold with $\alpha=1/1152$.

\paragraph{Transferring the error.}
The lower bound also makes $\Map$ injective on $\im L$, since
$L^{1/2}$ maps that space onto itself. Moreover, $\Map$ vanishes
on $\ker L$, so $\im\Map^\top=\im L$. Let $L'$ be a reweighted
Laplacian using edges of $L$. Both vanish on $\ker L$, so it
suffices to compare their energies on $\im L$. Every $f$ in this
space has the form $f=\Map^\top z$, and hence
\[
 f^\top L'f=z^\top(\Map L'\Map^\top)z,\qquad
 f^\top Lf=z^\top S_0z .
\]
The relative error guarantee for $\Map L'\Map^\top$ with respect
to $S_0$ therefore gives the same guarantee for $L'$ with respect to $L$.

\paragraph{Computing a probe on the graph.}
Fix a phase starting with $p$ edges of weights $w_e$ and let $C$
have rows $\sqrt{w_e}\,b_e^\top$ for those edges. The normalized
edge matrices are $B_e=w_e(\Map b_e)(\Map b_e)^\top$, and in the
first phase they sum to $S_0$ of rank at most $n$. For vectors $u,v$
in the normalized space, set $f=\Map^\top u$ and $g=\Map^\top v$. The entries
of $Cf$ and $Cg$ are the weighted endpoint differences, so the
bilinear forms and the matrix product in \Cref{thm:partial} are
\begin{equation}\label{eq:edgeops}
 u^\top B_e v=(Cf)_e(Cg)_e,\qquad
 H(x)u=\rho^{-1}\Map C^\top\diag(x)Cf .
\end{equation}
The map $\Map$ stays fixed through all coloring phases, $C$ uses the
weights at the start of the phase, and $\diag(x)$ changes at each step.
Applying $C$ or $C^\top$ takes $O(p+n)$ work. Applying $\Map=DW/8$
or its transpose takes one solver application and one pass over the
$O(n\log n)$ rows of $D$, for $O(n\log n)$ work. This row count is
also the dimension of $\xi$ and $v$, so drawing $\xi$ and computing
$\|v\|_2^2$ fit the same bound. Each operation in \eqref{eq:edgeops}
therefore takes $O(p+n\log n)$ work, including the scan that computes
all edge forms.

\Cref{alg:graph} summarizes the graph algorithm. All matrices are
represented through the operations in \eqref{eq:edgeops}, and any
reported failure terminates the algorithm.

\begin{algorithm}[htbp]
\caption{Spectral sparsification of $G$ with accuracy $\eps$}\label{alg:graph}
\begin{enumerate}[leftmargin=2.4em,itemsep=2pt,topsep=3pt]
\item Discard loops and combine parallel edges. Set
$\eta\gets\eps/4$ and $p_*\gets C_0n/\eta^2$.
\item If $|E(G)|\le p_*$, return $G$.
\item If $|E(G)|>n\log n/\eps^2$, use the sampler above with
$\delta=\eps/8$; otherwise keep $G$. Denote the resulting graph by $G_0$.
\item If $|E(G_0)|\le p_*$, return $G_0$.
\item Set $L\gets L_{G_0}$ and construct $\Map=DW/8$ as above.
Keep $\Map$ fixed. Choose even $\ell=\Theta(\log n)$ sufficiently large.
\item Set $m_0\gets|E(G_0)|$, $q\gets1-\gamma/2$,
$K\gets\min\{j\ge0:m_0q^j\le p_*\}$, $t\gets0$, and $G'\gets G_0$.
\item While $p_t=|E(G')|>p_*$, perform the following steps.
\begin{enumerate}[label=(\alph*),leftmargin=2em,itemsep=2pt,topsep=2pt]
\item Choose $\rho_t$ from \eqref{eq:schedule} with
$d=n$ and accuracy parameter $\eta$.
\item For each current edge, define
$B_e=w_e(\Map b_e)(\Map b_e)^\top$ implicitly.
\item Obtain $y$ from
\proc{PartialColor}$((B_e)_{e\in E(G')};\rho_t,\ell)$.
\item In $G'$, set $w_e\gets w_e(1+y_e)$ and remove the
zero-weight edges. Set $t\gets t+1$.
\end{enumerate}
\item Return $G'$.
\end{enumerate}
\end{algorithm}

\paragraph{Preprocessing time.}
On a graph with $r$ edges the solver construction takes
$O(r\log n+n\log^2 n)$ work with high probability by
\cite[full version, Lemma~I.5]{KLPSS} with constant recursion
parameter. The same bound holds for the initial sampling by
their Lemma~H.1. It constructs an approximate inverse on a sample,
applies it $O(\log n)$ times at $O(n\log n)$ work per application,
and spends $O(r\log n)$ work on the edge scores and sampling.

Cap the initial sampling and the solver construction at sufficiently
large constant multiples of their work and output-size bounds,
and report failure if a cap is exceeded. The high-probability
guarantees above imply that both calls meet these bounds and return
valid outputs with high probability. Thus preprocessing takes
$O(m\log n+n\log^2 n)$ time on every run.

\paragraph{Proof of \Cref{thm:main}.}
Take an even $\ell=\Theta(\log n)$ with constant large enough for
the desired failure exponent. Conditional on successful preprocessing
and all preceding colorings, \Cref{thm:partial} bounds the failure
probability in phase $t$ by $e^{-\ell}+e^{-p_t/64}$. We have
$p_t>16C_0n/\eps^2$ in each of at most $K=O(\log\log n)$ phases,
so a union bound gives success with high probability.

On success,
\Cref{sec:schedule} and the energy identities above give a
$(1\pm\eps/4)$ approximation to $L$ with at most $16C_0n/\eps^2$
edges. Including the initial sampling changes the lower factor
to $(1-\eps/4)e^{-\eps/8}\ge(1-\eps/4)(1-\eps/8)\ge1-\eps$ and
the upper factor to $(1+\eps/4)e^{\eps/8}\le(1+\eps/4)^2\le1+\eps$.

By \Cref{thm:partial} and \eqref{eq:edgeops}, a phase costs
$O(\rho_t^{-3}(p_t+n\log n)\log^2 n)$. Summing the edge and
solver costs by \eqref{eq:sum-work} gives total coloring time
\begin{equation}\label{eq:graph-work}
 O\!\left(\eps^{-3}\log^2 n\,\bigl[m_0+n\log n\,K^4\bigr]\right).
\end{equation}
The step limit within a call and the bound on the surviving edge
count hold on every run. Substituting $m_0\le m$ and
$K=O(\log\log n)$ and adding preprocessing proves the worst-case
time bound.

\section*{Acknowledgments}

The author thanks Jason Li for suggestions regarding the exposition of this paper and Richard Peng for his helpful feedback.

\section*{AI methodology}

The author presented an outline of the algorithm to GPT-6
Astra Pro, which assisted in developing some of the proofs and wrote
up the resulting work. The author then used the same model to remove
a factor of $1/\eps$ from the running time and simplify the algorithm.
The model also assisted in drafting and revising the exposition of
this paper.

\bibliographystyle{alpha}
\bibliography{references}
\end{document}